\documentclass{optica-article}

\journal{opticajournal} 

\articletype{Research Article}

\usepackage{lineno,amsmath,csquotes,graphicx,psfrag,bm,caption,float,xcolor}

\begin{document}

\title{Hanbury Brown--Twiss effect and classical entanglement with OAM-carrying light}

\author{Jyrki Laatikainen,\authormark{*} Sushil Pokharel, and Olga Korotkova}

\address{Department of Physics, University of Miami, 1320 Campo Sano Avenue, Coral Gables, FL, 33146, USA}

\email{\authormark{*}jjl376@miami.edu} 


\begin{abstract*} 
Intensity-intensity correlation of a scalar optical beam carrying orbital angular momentum (OAM) across multiple modes is decomposed into intermodal contributions, thereby linking it, within the framework of the Hanbury Brown--Twiss effect, to the underlying modal coherence structure. Upon filtering the spiral phase dependence, the intensity-intensity correlations are governed by OAM-mode coherence and anisotropy reflecting classical entanglement between OAM and statistical properties of light. These results extend intensity interferometry to structured light fields and provide direct access to modal coherence properties without phase-sensitive measurements.
\end{abstract*}

\section{Introduction}
The Nobel-prize nominated Hanbury Brown--Twiss (HBT) effect, first demonstrated in 1956, revealed that the intensity correlations can encode spatial information about a source even when amplitude or phase information is not measured \cite{HBT1,HBT2}. This finding inaugurated the field of intensity interferometry, where 
information about optical fields is extracted from intensity fluctuations, providing a powerful tool for studying light from thermal or chaotic sources. The quantum-optical framework developed by Glauber in 1963, later leading to a Nobel Prize, formalized this approach in terms of intensity correlation functions of higher orders, laying the foundation for photon statistics and quantum coherence phenomena \cite{Glauber1,Glauber2}. Subsequent studies connected intensity correlations with electromagnetic (EM) coherence and polarization \cite{Tervo2003,Tero2004,Shirai2007,Volkov2008,Hassinen2011,Shirai2017,TacoEM}, enabling applications from holography \cite{Singh2018} to beam propagation through optical systems \cite{Jacks1,TacoLens}, turbulence \cite{Korotkova2008,Ata}, and scattering media \cite{Jacks2}.

By contrast, the application of intensity interferometry to light carrying orbital angular momentum (OAM) remains comparatively unexplored, largely because the spiral phase structure of OAM-carrying modes \cite{Allen1992} is not directly encoded in intensity measurements. 
Particular exceptions are Refs.~\cite{Li2011,Singh2012,Boyd2016,Boyd2017,Dogariu2019}, which explore intensity interferometry with individual OAM modes and the related applications. Recent observations of annular correlation structures for modes with different OAM indices in scattering and turbulent media \cite{ScattStatic,ScattDynamic,Bastian,Rumman2026} further highlight the conservation of nontrivial intermodal correlations. However, a general framework for the HBT effect for light oscillating in multiple OAM modes remains missing.

In this work, we achieve a decomposition of the intensity–intensity correlation function of a scalar, statistically stationary light beam into contributions associated with those of the OAM modes comprising it. The decomposition is further examined for the correlation obtained after filtering out the azimuthal phase dependence of each OAM component. In this regime, the results previously known for two-component partially polarized EM fields \cite{Volkov2008,Hassinen2011} are generalized to optical fields containing an arbitrary number of OAM modes. Under Gaussian statistics assumption, the radially resolved intensity correlation function is governed by measures of modal coherence and anisotropy of the beam, which manifest classical entanglement in the OAM and spatial correlation degrees of freedom. These findings establish the HBT effect in OAM space, providing a framework for interpreting intensity correlations in structured light fields and accessing OAM-mode coherence using intensity interferometry.

\section{Principle of Hanbury Brown--Twiss effect}
\begin{figure}[t!]
    \centering
\includegraphics[width=0.75\columnwidth]{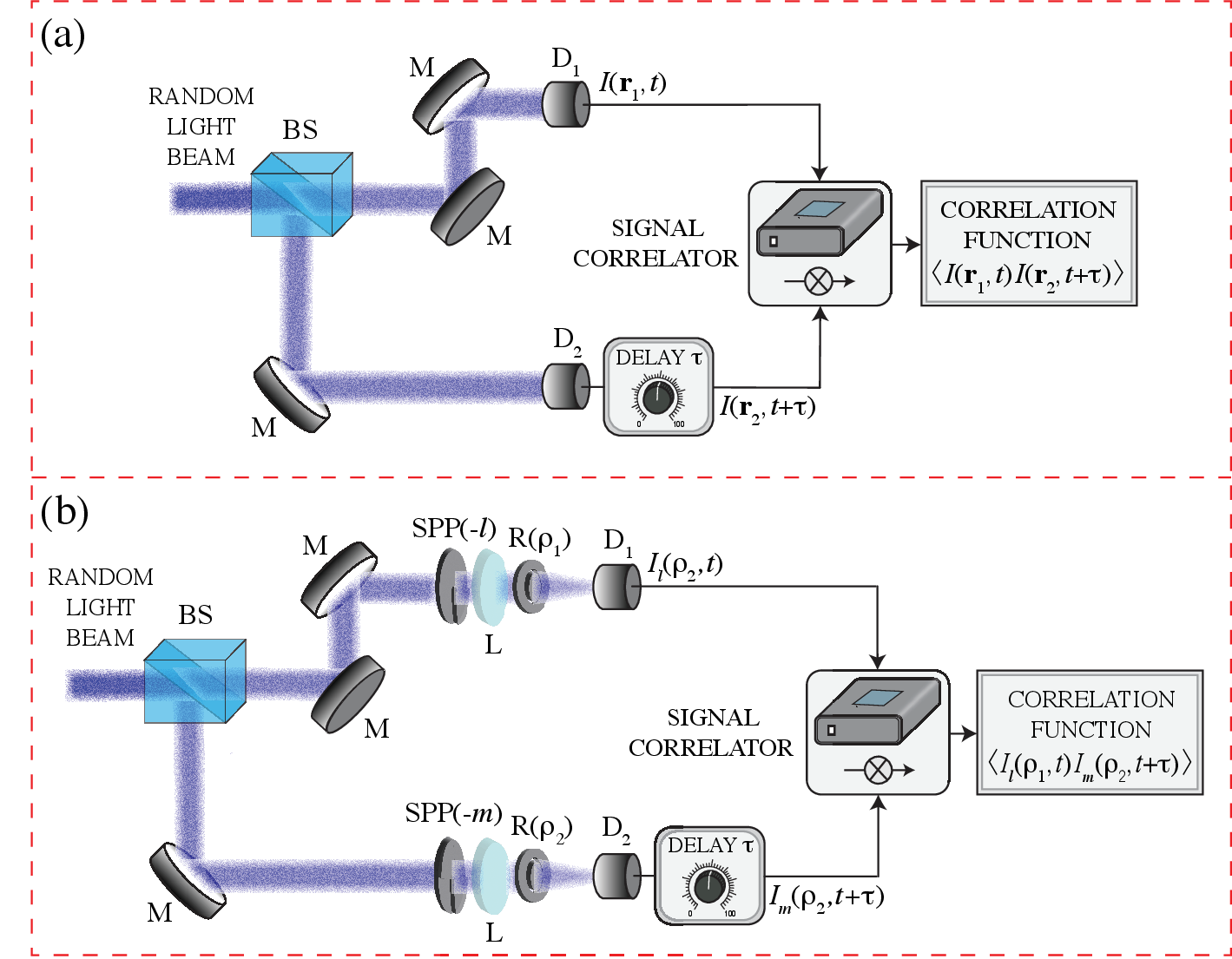} 
\caption{The HBT inteferometer for (a) the correlation $\langle I(\mathbf{r}_1,t)I(\mathbf{r}_2,t+\tau)\rangle$ and (b) the radially and OAM-resolved correlation $\langle I_l(\rho_1,t)I_m(\rho_2,t+\tau)\rangle$. A random beam is divided by a beam splitter (BS) and directed by mirrors (M) to detectors $\mathrm{D}_1$ and $\mathrm{D}_2$. The signal from $\mathrm{D}_2$ is delayed by $\tau$, after which the signals are correlated. In (b), the spiral phase plates (SPP) of orders $-l$ and $-m$, lenses (L), and ring apertures (R) of radii $\rho_n$ at a plane $z=0$ before $\mathrm{D}_1$ and $\mathrm{D}_2$ enable detection of intensities $I_l(\rho_1,t)$ and $I_m(\rho_2,t+\tau)$. 
}\label{fig:HBT}
\end{figure}

We begin by considering a scalar, statistically stationary, ergodic optical beam field entering the HBT interferometer depicted in Fig.~\ref{fig:HBT}(a). The HBT setup contains photodetectors $\mathrm{D}_1$ and $\mathrm{D}_2$ at points $\mathbf{r}_1$ and $\mathbf{r}_2$, respectively, which generate photocurrents proportional to the instantaneous intensity of light incident on them. The signal coming from $\mathrm{D}_2$ is delayed by a time $\tau$, after which the signals are correlated. If we denote with $U(\mathbf{r},t)$ the complex field at a position $\mathbf{r}$ and time $t$, the intensity fluctuation at $(\mathbf{r},t)$ is
\begin{align}\label{DeltaI}
     \Delta I(\mathbf{r},t)=I(\mathbf{r},t)-S(\mathbf{r}).
\end{align}
Above, $I(\mathbf{r},t)=\left|U(\mathbf{r},t)\right|^2$ is the instantaneous intensity of the beam, $S(\mathbf{r})=\langle I(\mathbf{r},t)\rangle$ is the mean intensity, and angle brackets represent time 
averaging. The correlation between the intensity fluctuations from the detectors, 
$G(\mathbf{r}_1,\mathbf{r}_2,\tau)=\langle \Delta I(\mathbf{r}_1,t) \Delta I(\mathbf{r}_2,t+\tau)\rangle$, is given by the (real-valued) covariance
\begin{align}\label{C12}
    G(\mathbf{r}_1,\mathbf{r}_2,\tau)=\langle I(\mathbf{r}_1,t)I(\mathbf{r}_2,t+\tau)\rangle-S(\mathbf{r}_1)S(\mathbf{r}_2),
\end{align}
which is customarily normalized as
\begin{align}\label{c12}
g(\mathbf{r}_1,\mathbf{r}_2,\tau)=\frac{G(\mathbf{r}_1,\mathbf{r}_2,\tau)}{S(\mathbf{r}_1)S(\mathbf{r}_2)}.
\end{align}
At equal spatial arguments $\mathcal{C}(\mathbf{r})=G(\mathbf{r},\mathbf{r},0)$ describes the scintillation, i.e., the intensity variance, of the field at $\mathbf{r}$, while its normalized version, $c(\mathbf{r})=g(\mathbf{r},\mathbf{r},0)$, is the scintillation index \cite{Korotkova2008}. 

An important class of light fields, including beams generated by many practical light sources, are those governed by complex Gaussian statistics. 
For such fields, the following moment theorem holds \cite{Mandel}:
\begin{align}\label{Gauss}
    \langle\mathrm{perm} (\vec{U}_{12}^\dagger \vec{U}_{34}) \rangle
    =2\,\mathrm{perm}\langle \vec{U}_{12}^\dagger \vec{U}_{34} \rangle,
\end{align}
where $\vec{U}_{ij}=[U(\mathbf{r}_i,t_i),U(\mathbf{r}_j,t_j)]$, $i,j\in\{1,2,3,4\}$, while perm and dagger denote the permanent and complex transpose, respectively. Relation \eqref{Gauss} implies that the information of the fourth-order correlation function is completely encoded in the second-order correlations.
In this case, Eq.~(\ref{c12}) leads to the classic result \cite{Mandel} 
\begin{align}\label{c12_gauss}
    g(\mathbf{r}_1,\mathbf{r}_2,\tau)=|\gamma (\mathbf{r}_1,\mathbf{r}_2,\tau)|^2,
\end{align}
where 
\begin{align}\label{gamma}
\gamma (\mathbf{r}_1,\mathbf{r}_2,\tau)=\frac{\Gamma(\mathbf{r}_1,\mathbf{r}_2,\tau)}{\sqrt{S(\mathbf{r}_1)S(\mathbf{r}_2)}}
\end{align}
is the complex degree of coherence 
and $\Gamma(\mathbf{r}_1,\mathbf{r}_2,\tau)=\langle U^\ast (\mathbf{r}_1,t)U(\mathbf{r}_2,t+\tau)\rangle$ is the mutual coherence function (MCF) 
describing correlation between the fields from the detectors, with asterisk denoting the complex conjugate. The degree of coherence satisfies $0\leq |\gamma(\mathbf{r}_1,\mathbf{r}_2,\tau)|\leq 1$, where the lower and upper bounds correspond, respectively, to a complete absence of second-order correlation and to perfect correlation between the fields at $\mathbf{r}_1$ and $\mathbf{r}_2$. The expression \eqref{c12_gauss} manifests that, for light with complex Gaussian statistics, the correlation of intensity fluctuations in the HBT experiment is equivalent to the second-order field correlation. Further, under these assumptions the scintillation index is unity, $c(\mathbf{r})=1$. 

\section{Intensity-intensity correlations via the OAM-mode representation of the field}
New insight into the results above is gained via the OAM-mode representation
of the field. 
By expressing the position vector 
$\mathbf{r}$ in cylindrical coordinates, $\mathbf{r}=\rho\hat{\rho}+\phi\hat{\phi}+z\hat{z}$, we may decompose the instantaneous field into its polar Fourier spectrum \cite{Gori2001} 
\begin{equation}\label{Udec1}
U(\mathbf{r}, t)=\sum\limits_{l=-\infty}^{\infty} U_l(\pmb{\xi},t)e^{il\phi}.
\end{equation} 
Above, $\pmb{\xi}=\rho\hat{\rho}+z\hat{z}$ 
and
the polar Fourier components 
\begin{equation}\label{Udec2}
U_l(\pmb{\xi},t)=\frac{1}{2\pi}\int\limits_{0}^{2\pi}U(\textbf{r}, t)e^{-il\phi}d\phi
\end{equation} 
can be interpreted as the complex amplitudes of the vortex modes $U_l(\mathbf{r},t)=U_l(\pmb{\xi},t)e^{il\phi}$, each carrying
OAM of $l\hbar$ per photon \cite{Allen1992}. As is relevant for most practical purposes, we will restrict the following analysis to situations considering a finite number of modes in Eq.~(\ref{Udec1}), which constitute a finite set $\mathcal{L}$ such that the dimension of the space spanned by them is $L$.

Employing Eq.~(\ref{Udec1}), we may express the correlation of the intensity fluctuations, Eq.~(\ref{c12}), as 
\begin{align}\label{G12-mat}
    G(\mathbf{r}_1,\mathbf{r}_2,\tau)=\langle\!\langle \overleftrightarrow{G}(\mathbf{r}_1,\mathbf{r}_2,\tau)\rangle\!\rangle_{L^2},
\end{align}
where $\langle\!\langle \overleftrightarrow{A} \rangle\!\rangle_\alpha=\vec{1}_{\alpha}^T\overleftrightarrow{A}\vec{1}_{\alpha}$ returns the sum of the elements in the $\alpha\times\alpha$ matrix $\overleftrightarrow{A}$, with $\vec{1}_{\alpha}$ representing an $\alpha$D vector of ones. In other words, $G(\mathbf{r}_1,\mathbf{r}_2,\tau)$ is obtained by summing over the elements of the $L^2\times L^2$ matrix $\overleftrightarrow{G}(\mathbf{r}_1,\mathbf{r}_2,\tau)=[G_{lmpq}(\mathbf{r}_1,\mathbf{r}_2,\tau)]$, with $l,m,p,q \in \mathcal{L}$ and
\begin{align}\label{Glmpq}
G_{lmpq}(\mathbf{r}_1,\mathbf{r}_2,\tau)&=\langle \Delta I_{lm}(\mathbf{r}_1,t) \Delta I_{pq}(\mathbf{r}_2,t+\tau)\rangle.
\end{align}
Above, $I_{lm}(\mathbf{r},t)=U_l^\ast(\mathbf{r},t)U_m(\mathbf{r},t)$  and $\Delta I_{lm}(\mathbf{r},t)=I_{lm}(\mathbf{r},t)-\langle I_{lm}(\mathbf{r},t)\rangle$. Further, for fields obeying 
Gaussian statistics we obtain, by combining Eqs.~(\ref{Gauss}) and (\ref{Udec2}):
\begin{align}\label{Gauss-radial}
    \langle\mathrm{perm}[(\vec{U}_{12}^{lm})^\dagger \vec{U}_{34}^{pq}] \rangle =2\;\mathrm{perm}\langle  (\vec{U}_{12}^{lm})^\dagger \vec{U}_{34}^{pq} \rangle,
\end{align}
with 
$\vec{U}_{ij}^{\alpha\beta}=[U_\alpha(\pmb{\xi}_i,t_i),U_\beta(\pmb{\xi}_j,t_j)]$. 
Note that since 
field $U(\textbf{r},t)$ is wide-sense stationary and complex Gaussian, $U_l(\pmb{\xi},t)$ in Eq. (8) are the result of a linear functional and, therefore, remain complex Gaussian for all $l$. In addition, the members of ensemble $\{ U_{l}(\pmb{\xi},t) \}$ remain jointly complex Gaussian \cite{Goodman}.
Employing Eq.~\eqref{Gauss-radial} 
 together with Eq.~(\ref{G12-mat}) 
yields
\begin{align}\label{G-gauss}
    G(\mathbf{r}_1,\mathbf{r}_2,\tau)=\left|\langle\!\langle\overleftrightarrow{D}_1^\dagger \overleftrightarrow{\Gamma}(\pmb{\xi}_1,\pmb{\xi}_2,\tau)\overleftrightarrow{D}_2\rangle\!\rangle_L\right|^2,
\end{align}
where 
$\overleftrightarrow{D}_n=\mathrm{diag}(e^{il\phi_n})$, $n\in\{1,2\}$ 
In addition, 
\begin{align}
\overleftrightarrow{\Gamma}(\pmb{\xi}_1,\pmb{\xi}_2,\tau)=[\Gamma_{lm}(\pmb{\xi}_1,\pmb{\xi}_2,\tau)]     
\end{align}
is the space-time domain $L\times L$ Coherence-OAM (COAM) matrix \cite{Olga2021} containing as its elements all the 
correlations $\Gamma_{lm}(\pmb{\xi}_1,\pmb{\xi}_2,\tau)=\langle U_l^\ast(\pmb{\xi}_1,t)U_m(\pmb{\xi}_2,t+\tau)\rangle$ between the OAM modes in the field. The COAM matrix shares a Wiener--Khintchine relation \cite{Mandel} with its space-frequency counterpart (see Supplement 1 for details).

Expressions (\ref{G12-mat}) and (\ref{G-gauss}) constitute the first main contribution of this work. We have expanded the scalar correlation $G(\mathbf{r}_1,\mathbf{r}_2,\tau)$ into a correlation matrix with elements $G_{lmpq}(\mathbf{r}_1,\mathbf{r}_2,\tau)$, which generally represent the cross-covariances between the components $I_{lm}(\mathbf{r},t)$ associated with the corresponding OAM modes at points $\mathbf{r}_1$ and $\mathbf{r}_2$. At $l=m$, these components are the intensities of the modes $U_l(\mathbf{r},t)$, while for $l\neq m$ they represent complex-valued cross-terms between the corresponding modes. Furthermore, Eq.~(\ref{G-gauss}) shows that the Gaussian intensity-intensity correlation can be expressed as a decomposition into terms linked to the COAM matrix. This suggests that the local intensity correlation function of a beam can be controlled by adjusting the radially dependent field correlations of OAM-mode pairs, therefore enabling tailoring of light fields with specific higher-order correlation structures. 
Similar results hold naturally also for the normalized correlation $g(\mathbf{r}_1,\mathbf{r}_2,\tau)$, as well as the scintillation $\mathcal{C}(\mathbf{r})$ and scintillation index $c(\mathbf{r})$. 
\section{Radially resolved intensity-intensity correlations}
We will now focus our attention on the correlations among the radially resolved intensities to quantify the related OAM-mode mixing in characterization of classical entanglement \cite{Spreew1998,Kim2002,Qian2011} in spatial and OAM degrees of freedom. We define the radially resolved intensity as 
\begin{align}\label{Ibar}
    \bar{I}(\pmb{\xi},t)=\frac{1}{2\pi}\int_0^{2\pi}I(\mathbf{r},t)d\phi=\sum_{l=-\infty}^\infty I_l(\pmb{\xi},t),
\end{align}
where $I_l(\pmb{\xi},t)=|U_l(\mathbf{r},t)|^2$, and identity $\int_0^{2\pi} e^{i\alpha\phi}d\phi=2\pi\delta_{\alpha 0}$ together with Eq.~(\ref{Udec1}) has been used to obtain the second equality, with $\delta_{\alpha\beta}$ representing the Kronecker delta. We see from the second equality in Eq.~(\ref{Ibar}) that $\bar{I}(\pmb{\xi},t)$ is the sum of the intensities of modes $U_l(\mathbf{r},t)$, which excludes the contributions of the cross-terms between modes $l\neq m$ in $I(\mathbf{r},t)$. Consequently, the radially resolved intensity can be synthesized as depicted in Fig.~\ref{fig:HBT}(b): the HBT setup is modified by a ring aperture, a lens, and a spiral phase plate placed before each detector, enabling measurement of individual OAM-mode intensities for a paraxial, quasi-monochromatic beam (see Supplement 1 for details). This approach is analogous to that described in \cite{Olga2021} and experimentally verified in \cite{COAMmeasurement} for measuring the COAM matrix. 

The fluctuation of the intensity $\bar{I}(\pmb{\xi},t)$ around its mean value $\bar{S}(\pmb{\xi})=\langle\bar{I}(\pmb{\xi},t)\rangle$ is defined as $\Delta\bar{I}(\pmb{\xi},t)=\bar{I}(\pmb{\xi},t)-\bar{S}(\pmb{\xi})$. The associated correlation $\bar{G}(\pmb{\xi}_1,\pmb{\xi}_2,\tau)=\langle\Delta\bar{I}(\pmb{\xi}_1,t)\Delta \bar{I}(\pmb{\xi}_2,t+\tau)\rangle$ is given by the cross-covariance
\begin{align}\label{C12-radial}
    \bar{G}(\pmb{\xi}_1,\pmb{\xi}_2,\tau)&=\langle \bar{I}(\pmb{\xi}_1,t)\bar{I}(\pmb{\xi}_2,t+\tau)\rangle - \bar{S}(\pmb{\xi}_1)\bar{S}(\pmb{\xi}_2)
     =\langle\!\langle \overleftrightarrow{M}(\pmb{\xi}_1,\pmb{\xi}_2,\tau)-\vec{S}(\pmb{\xi}_1)\vec{S}^\mathrm{T}(\pmb{\xi}_2) \rangle\!\rangle_L,
 \end{align}
which is further normalized as 
\begin{align}\label{c12-radial}
    \bar{g}(\pmb{\xi}_1,\pmb{\xi}_2,\tau)&=\frac{\bar{G}(\pmb{\xi}_1,\pmb{\xi}_2,\tau)}{\bar{S}(\pmb{\xi}_1) \bar{S}(\pmb{\xi}_2)}.
\end{align}
Here 
$\vec{S}(\pmb{\xi})=[\langle I_l(\pmb{\xi},t)\rangle ]$ is a $L$D vector containing the mean intensities of the OAM modes in the field, while $\overleftrightarrow{M}(\pmb{\xi}_1,\pmb{\xi}_2,\tau)=[\langle  I_l(\pmb{\xi}_1,t)I_m(\pmb{\xi}_2,t+\tau) \rangle ]$ is a $L\times L$ matrix of corresponding modal intensity-intensity correlations. 

For fields governed by complex Gaussian statistics, \(\bar{I}(\pmb{\xi},t) \) is a sum of squared Gaussian amplitudes and therefore not itself Gaussian. Nevertheless, because the field $U(\textbf{r},t)$ is complex Gaussian, all moments of \(\bar{I}(\pmb{\xi},t)\), including the intensity-intensity correlation, can be computed using the Gaussian moment theorem, Eq.~(\ref{Gauss-radial}), applied to the field amplitudes \cite{Goodman}. In this case, 
\begin{align}
&\overleftrightarrow{M}(\pmb{\xi}_1,\pmb{\xi}_2,\tau)-\vec{S}(\pmb{\xi}_1)\vec{S}^\mathrm{T}(\pmb{\xi}_2)
=\overleftrightarrow{\Gamma}(\pmb{\xi}_1,\pmb{\xi}_2,\tau)\odot\overleftrightarrow{\Gamma}^\ast(\pmb{\xi}_1,\pmb{\xi}_2,\tau),    
\end{align}
where $\odot$ denotes the Hadamard product. Combining this relation with Eq.~(\ref{c12-radial}) leads to
\begin{align}\label{cbar-gauss}
    \bar{g}(\pmb{\xi}_1,\pmb{\xi}_2,\tau)=\bar{\gamma}^2(\pmb{\xi}_1,\pmb{\xi}_2,\tau),
\end{align}
where 
\begin{align}\label{OAMDC}
\bar{\gamma}(\pmb{\xi}_1,\pmb{\xi}_2,\tau)=\sqrt{\frac{\mathrm{tr}[\overleftrightarrow{\Gamma}^\dagger(\pmb{\xi}_1,\pmb{\xi}_2,\tau)\overleftrightarrow{\Gamma}(\pmb{\xi}_1,\pmb{\xi}_2,\tau)]}{\bar{S}(\pmb{\xi}_1) \bar{S}(\pmb{\xi}_2)}}
\end{align}
is the space-time domain OAM degree of coherence (OAMDC) \cite{Olga2024} characterizing the total amount of correlation within the OAM-mode distribution of the field, with tr denoting the trace. 
We see directly from the Schwarz inequality $|\langle U_l^\ast(\pmb{\xi}_1,t) U_m(\pmb{\xi}_2,t+\tau) \rangle|^2\leq \langle |U_l(\pmb{\xi}_1,t)|^2\rangle\langle |U_m(\pmb{\xi}_2,t)|^2\rangle$ that the OAMDC satisfies $0\leq \bar{\gamma}(\pmb{\xi}_1,\pmb{\xi}_2,\tau)\leq 1$. As shown in Supplement 1, the lower and upper limits imply that the OAM-components within the field at $\pmb{\xi}_1$ and $\pmb{\xi}_2$ are fully correlated and uncorrelated, respectively, for all pairs $l$ and $m$ at the time delay $\tau$. 

Expression~(\ref{cbar-gauss}), being the second major result of this work, is analogous in form to that obtained for EM beam fields obeying Gaussian statistics in the HBT experiment \cite{Hassinen2011}, as well as the scalar-field result displayed in Eq.~(\ref{c12_gauss}) to which Eq.~(\ref{cbar-gauss}) reduces for a single OAM mode. It shows that all the information of the radially resolved Gaussian intensity-intensity correlation is contained in the OAMDC of the field. Alternatively, it implies that the OAMDC can be accessed solely via intensity correlation measurements, without involving direct second-order field correlation measurements.

Further insight into Eq.~(\ref{cbar-gauss}) is gained via the $L^2$ 
Stokes parameters related to the COAM matrix \cite{Olga2021OL}:
\begin{align}
    \bar{\mathcal{S}}_n(\pmb{\xi}_1,\pmb{\xi}_2,\tau)=\mathrm{tr}[\overleftrightarrow{\Lambda}_n\overleftrightarrow{\Gamma}(\pmb{\xi}_1,\pmb{\xi}_2,\tau)], 
\end{align}
where $n\in\{0,\dots,L^2-1\}$, $\overleftrightarrow{\Lambda}_0$ is the $L$D identity matrix and $\overleftrightarrow{\Lambda}_n$, $n\geq 1$, are the generalized Gell--Mann matrices \cite{GM}. The parameters $\bar{\mathcal{S}}_n(\pmb{\xi}_1,\pmb{\xi}_2,\tau)$ are complex-valued and can be normalized as
\begin{align}
\bar{\gamma}_n(\pmb{\xi}_1,\pmb{\xi}_2,\tau)=\frac{\bar{\mathcal{S}}_n(\pmb{\xi}_1,\pmb{\xi}_2,\tau)}{\sqrt{\bar{\mathcal{S}}_0(\pmb{\xi}_1,\pmb{\xi}_1,0)\bar{\mathcal{S}}_0(\pmb{\xi}_2,\pmb{\xi}_2,0)}},
\end{align}
Noting that $\mathrm{tr}(\overleftrightarrow{\Lambda}_n\overleftrightarrow{\Lambda}_k)=2\delta_{nk}$, $n,k>0$, the OAMDC can be written as
\begin{align}
    \bar{\gamma}^2(\pmb{\xi}_1,\pmb{\xi}_2,\tau)=\frac{1}{L}|\bar{\gamma}_0(\pmb{\xi}_1,\pmb{\xi}_2,\tau)|^2+\frac{1}{2}\sum_{n=1}^{L^2-1}|\bar{\gamma}_n(\pmb{\xi}_1,\pmb{\xi}_2,\tau)|^2.
\end{align}
This expression is analogous to that between the 2D EM degree of coherence and the coherence Stokes parameters \cite{Friberg2016}. One may then rewrite Eq.~(\ref{cbar-gauss}) (see Supplement~1):
\begin{align}\label{23}
    \bar{g}(\pmb{\xi}_1,\pmb{\xi}_2,\tau)=\frac{1}{L}\left[1+(L-1)\mathcal{Q}^2(\pmb{\xi}_1,\pmb{\xi}_2,\tau)\right]|\bar{\gamma}_0(\pmb{\xi}_1,\pmb{\xi}_2,\tau)|^2,
\end{align}
where 
\begin{align}
    &\mathcal{Q}(\pmb{\xi}_1,\pmb{\xi}_2,\tau)
    =\sqrt{\frac{L}{2(L-1)}\frac{\sum_{n=1}^{L^2-1}|\bar{\mathcal{S}}_n(\pmb{\xi}_1,\pmb{\xi}_2,\tau)|^2}{|\bar{\mathcal{S}}_0(\pmb{\xi}_1,\pmb{\xi}_2,\tau)|^2}}
\end{align}
is defined analogously to the degree of cross-polarization of 2D EM beams \cite{Volkov2008}. Similarly to it, $\mathcal{Q}(\pmb{\xi}_1,\pmb{\xi}_2,\tau)$ is a real-valued, non-negative, and non-limited quantity. Setting $\pmb{\xi}_1=\pmb{\xi}_2$, we find that
\begin{align}
    \bar{g}(\pmb{\xi})=\bar{\gamma}^2(\pmb{\xi},\pmb{\xi},0)=\frac{1}{L}\left[1+(L-1)Q^2(\pmb{\xi})\right].
\end{align}
Here $Q(\pmb{\xi})=\mathcal{Q}(\pmb{\xi},\pmb{\xi},0)$, which, by the orthogonality of $\overleftrightarrow{\Lambda}_n$, $n\geq 1$, can also be expressed as 
\begin{align}
    Q(\pmb{\xi})=\sqrt{\frac{L}{L-1}\left\{\frac{\mathrm{tr}[\overleftrightarrow{O}^2(\pmb{\xi})]}{\mathrm{tr}^2[\overleftrightarrow{O}(\pmb{\xi})]}-\frac{1}{L}\right\}}.
\end{align}
Above, $\overleftrightarrow{O}(\pmb{\xi})=\overleftrightarrow{\Gamma}(\pmb{\xi},\pmb{\xi},0)$ is the space-time domain orbitalization matrix (OM) \cite{OErandom} associated with the scalar random field, which is Hermitian and non-negative definite. If we denote the (non-negative) eigenvalues of $\overleftrightarrow{O}(\pmb{\xi})$ as $\lambda_1(\pmb{\xi})\geq \dots\geq \lambda_L(\pmb{\xi})$, the relations 
\begin{align}
\mathrm{tr}[\overleftrightarrow{O}(\pmb{\xi})] = \sum_{l=1}^{L} \lambda_l(\pmb{\xi}), \quad \mathrm{tr}[\overleftrightarrow{O}^2(\pmb{\xi})] = \sum_{l=1}^{L} \lambda_l^2(\pmb{\xi}),
\end{align}
yield 
\begin{align}\label{Qeig}
Q(\pmb{\xi}) = \sqrt{\frac{L}{L-1} \left\{ \sum_{l=1}^{L} \sigma_l^2(\pmb{\xi}) -\frac{1}{L}\right\}},
\end{align}
with $\sigma_l(\pmb{\xi})=\lambda_l(\pmb{\xi})/\sum_{l=1}^{L} \lambda_l(\pmb{\xi})$. We see from $1/L\leq \sum_{l=1}^{L} \sigma_l^2(\pmb{\xi})\leq 1$ that $Q(\pmb{\xi})$ is bound as $0\leq Q(\pmb{\xi})\leq 1$. The upper bound is achieved when all the power is in $\lambda_1(\pmb{\xi})$, i.e., $\lambda_l(\pmb{\xi})=0$, $l>1$, while at the lower bound all the eigenvalues are equal, such that  $\sigma_l(\pmb{\xi})=1/L$. Consequently, the situations $Q(\pmb{\xi})=1$ and $Q(\pmb{\xi})=0$ represent
completely orbitalized and unorbitalized states \cite{OErandom}, respectively, as classified by the space-time domain OM. In the former case the OM factors, while the latter accounts for all its spectral components carrying the same weight and thus none dominating the others.

Measure $Q(\pmb{\xi})$ appears to be properly defining mode mixing in $L$D \cite{Samson1973},  obeying the required majorization constraints \cite{James2014}. It is also the first in the hierarchy of purity measures according to Barakat's classification \cite{Barakat1983}. Note that $Q(\pmb{\xi})$ can also be defined via the spectral distance of the OM from a uniform distribution (see Supplement 1 for details) and, hence, be regarded as the degree of orbital anisotropy (DOA) of a scalar field in the OAM space.

\begin{table}
\centering
\begin{tabular}{l p{0.55\textwidth} l}
\hline\hline
Quantity & \quad \quad \quad Definition & Limits \\
\hline

Schmidt number &
$
\begin{aligned}[t]
\quad \quad \quad K(\pmb{\xi})
 &=\textstyle 1/\sum_{l=1}^L \sigma_l^2(\pmb{\xi}) \\
 &=\textstyle L/[1+(L-1)Q^2(\pmb{\xi})]
\end{aligned}
$
&
$1 \le K(\pmb{\xi}) \le L$
\\[3em]

Concurrence &
$
\begin{aligned}[t]
\quad \quad \quad C(\pmb{\xi})
 &=\textstyle \sqrt{L[1-\sum_{l=1}^L \sigma_l^2(\pmb{\xi})]/(L-1)} \\
 &= \textstyle\sqrt{1-Q^2(\pmb{\xi})}
\end{aligned}
$
&
$0 \le C(\pmb{\xi}) \le 1$
\\[3em]

Shannon's entropy &
$
\begin{aligned}[t]
\quad \quad \quad H(\pmb{\xi})
 &\textstyle= \sum_{l=1}^{L}\sigma_l(\pmb{\xi})
    \ln[\sigma_l(\pmb{\xi})] \\
 &\approx
 \ln\{L/[1+(L-1)Q^2(\pmb{\xi})]\}
\end{aligned}
$
&
$0 \le H(\pmb{\xi}) \le \ln L$
\\

\hline\hline
\end{tabular}
\caption{Relations between the DOA and measures of classical entanglement related to coupling between the spatial correlation and OAM degrees of freedom.}
\label{table1}
\end{table}

The DOA is closely related to measures characterizing classical entanglement, i.e., non-separable correlations within internal degrees of freedom of a light wave \cite{Spreew1998,Kim2002,Qian2011}. Classical entanglement between OAM and statistical properties is present in the OM as long as at least two of its eigenvalues $\lambda_{l}(\pmb{\xi})$ are non-trivial, implying that the matrix does not factor and the OAM and (radial) spatial correlation characteristics of the field are thus non-separable \cite{OErandom}. Table~\ref{table1} summarizes relations existing between the DOA and several widely used classical entanglement measures. In particular, radial Schmidt number $K(\boldsymbol{\xi})$ \cite{Qian2011,Grobe1994} quantifies the effective number of participating OAM modes, ranging from unity for a pure (completely orbitalized) state to $L$ for a maximally entangled (completely unorbitalized) state. Radial concurrence $C(\boldsymbol{\xi})$ \cite{McLaren2015,Peters2023}, the vector quality factor, measures the degree of coupling between statistical and OAM degrees of freedom in the limits from zero to unity for the two aforementioned states. Finally, radial Shannon's entropy $H(\boldsymbol{\xi})$ \cite{Shannon,Refregier2004} quantifies the total disorder of the reduced OAM state, taking on limiting values of zero and $\ln L$ for the corresponding states. While the first two measures are exactly related to the DOA, the third is in an approximate relation with it, for sufficiently large $L$. The expression for the DOA derived from the radial intensity-intensity correlations, Eq. (\ref{Qeig}), and its relations to the radial classical entanglement measures summarized in Table \ref{table1} constitute the third major result of the work. 

\section{Numerical demonstration}

\begin{figure}[t]
    \centering
\includegraphics[width=0.75\columnwidth]{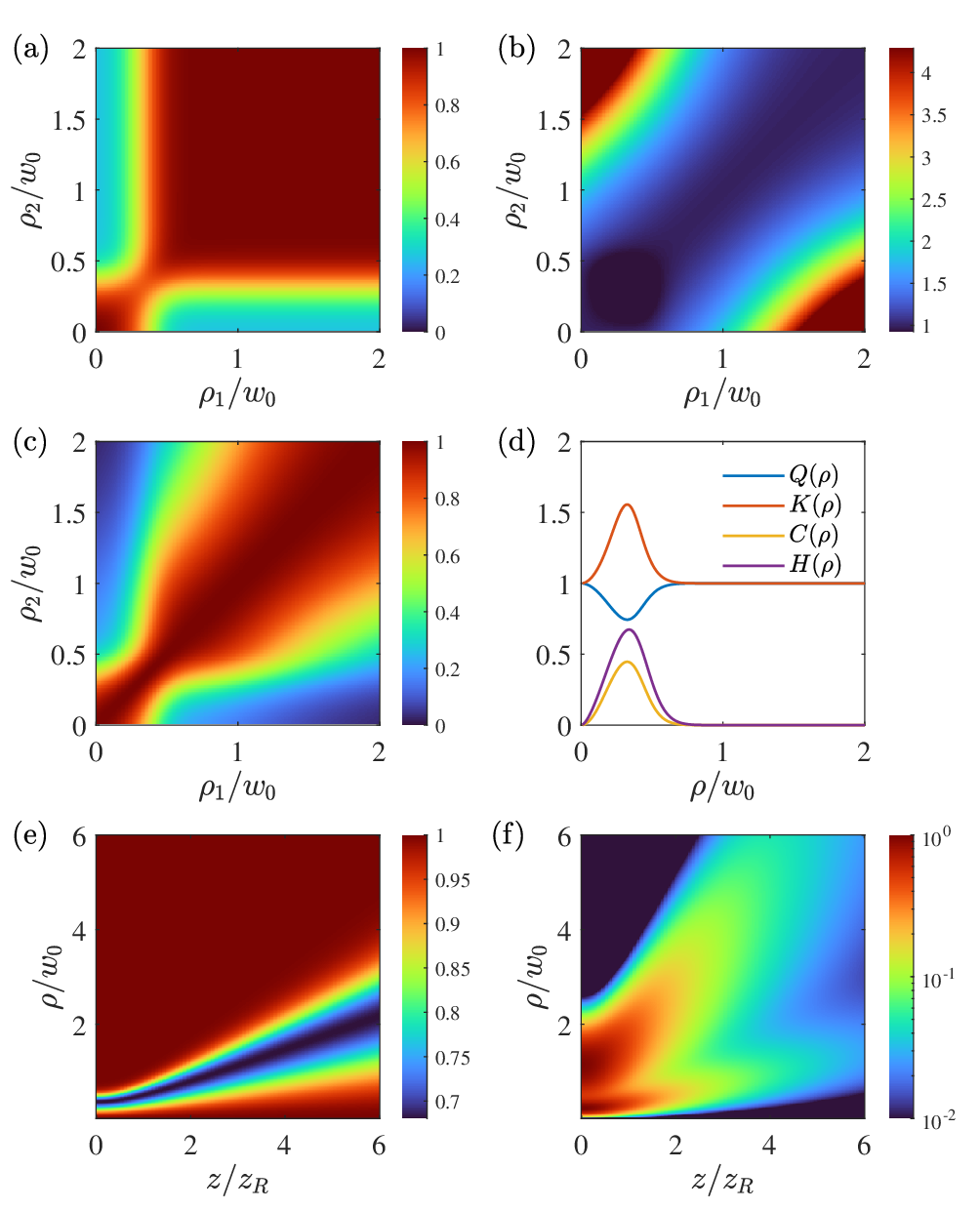}
    \caption{Illustration of (a) $\bar{\gamma}(\rho_1,\rho_2,0)$, (b) $\mathcal{Q}(\rho_1,\rho_2,0)$, (c) $|\gamma_0(\rho_1,\rho_2,0)|$, and (d) $Q(\rho)$, $K(\rho)$, $C(\rho)$, and $H(\rho)$. Panels (e) and (f) show the evolution of $Q(\pmb{\xi})$ and $\bar{S}(\pmb{\xi})$, normalized to its maximum value, along $\rho$ and $z$. The results are obtained for a beam formed as a superposition of LG and $I_l$-Bessel correlated modes, with COAM matrix given in Eq.~(\ref{COAM-example}).
    } 
    \label{fig:example}
\end{figure}

We will now demonstrate the derived quantities with a numerical model of a beam composed as an incoherent superposition of monochromatic Laguerre-Gaussian (LG) \cite{Allen1992} and quasi-monochromatic, partially coherent $I_l$-Bessel correlated \cite{Ponomarenko2001} modes. The field correlations are assumed to follow the Gaussian moment theorem given by Eq.~(\ref{Gauss}). In addition, we set $\tau=0$ to focus on the spatial coherence characteristics. The beam's COAM matrix at a plane transverse to propagation direction $z$ is (see Supplement 1 for details)
\begin{align}\label{COAM-example}
    \overleftrightarrow{\Gamma}(\rho_1,\rho_2,z)=\vec{U}^\dagger(\rho_1,z)\vec{U}(\rho_2,z)+\overleftrightarrow{R}(\rho_1,\rho_2,z),
\end{align}
where components of $\vec{U}(\rho,z)=[U_l(\rho,z)]$ are LG modes with zeroth radial order and helical index $l$, and $\overleftrightarrow{R}(\rho_1,\rho_2,z)=\mathrm{diag}[R_l(\rho_1,\rho_2,z)]$ contains the contribution of the $I_l$-Bessel correlated modes. For the numerical evaluation, we consider modes $l\in\{1,2,3,4,5\}$, with beam waist $w_0=1$~mm and central wavelength $\lambda=632$~nm. The coherence parameter of the $I_l$-Bessel correlated modes is chosen as $\zeta=0.75$, corresponding to a moderately incoherent regime.

Figures~\ref{fig:example}(a)-(c) present the quantities $\bar{\gamma}(\rho_1,\rho_2,0)$, $\mathcal{Q}(\rho_1,\rho_2,0)$, and $|\gamma_0(\rho_1,\rho_2,0)|$ obtained from Eq.~(\ref{COAM-example}) at the waist plane. The OAMDC shown in Fig.~\ref{fig:example}(a), which incorporates the correlation $\bar{g}(\rho_1,\rho_2,0)$, exhibits overall behavior similar to the zeroth-order Stokes parameter in Fig.~\ref{fig:example}(c). Deviations arise in symmetric regions defined by $0.5<\rho_1/w_0<1$, $1<\rho_2/w_0<2$ and $1<\rho_1/w_0<2$, $0.5<\rho_2/w_0<1$, reflecting the contribution of the off-diagonals of the COAM matrix. The function $\mathcal{Q}(\rho_1,\rho_2,0)$ acts as a compensating factor in Eq.~(\ref{23}) when $|\gamma_0(\rho_1,\rho_2,0)|$ becomes small. In addition, Fig.~\ref{fig:example}(d) shows the corresponding single-radius function $Q(\rho)$, and also $K(\rho)$, $C(\rho)$, and $H(\rho)$. The single-radius quantities exhibit nontrivial structure primarily for $\rho<w_0$, indicating the region where the random $I_l$-Bessel component is most significant, producing maximal OAM-mode mixing. In this regime, $K(\rho)$, $C(\rho)$, and $H(\rho)$ display the expected inverse behavior relative to the DOA. Finally, Figs.~\ref{fig:example}(e) and (f) show the evolution of the DOA and the normalized form of the mean radial intensity $\bar{S}(\pmb{\xi})$ along $\rho$ and $z$. We see that the DOA of this beam model remains structurally invariant in free-space propagation and only radially expands due to diffraction, in analogy with other orbitalization metrics \cite{SushilOE2026}. 

\section{Conclusions}

In summary, we decomposed the intensity-intensity correlation function of a structured, random optical field into contributions pertaining to the OAM modes in the field. We further showed that, upon filtering the azimuthal phase dependence, the resulting correlation function admits a form directly analogous to that known for random EM beams, and, for fields governed by Gaussian statistics, is connected to classical entanglement in their statistical and OAM structures. The results establish a framework of the HBT effect for multimode OAM-carrying light. They enable analysis of intensity interferometry in structured light fields and provide access to modal coherence properties using intensity-only measurements, with potential applications in communication, sensing, and imaging systems employing structured light.

\begin{backmatter}
\bmsection{Funding}
This work was supported by the Finnish Cultural Foundation.


\bmsection{Disclosures}
The authors declare no conflicts of interest.

\bmsection{Data Availability Statement}
Data underlying the results presented in this paper are not publicly available at this time but may be obtained from the authors upon reasonable request.

\bmsection{Supplemental document}
See Supplement 1 for supporting content.

\end{backmatter}

\end{document}


\maketitle

\section{Relation between the space-time and space-frequency domain COAM matrices}\label{appA}
Here we derive the relation between the space-time domain COAM matrix $\overleftrightarrow{\Gamma}(\pmb{\xi}_1,\pmb{\xi}_2,\tau)$ and its space-frequency domain counterpart. By using Eqs.~(7) and (8), the following relations are found  between the MCF and COAM matrix elements:
\begin{align}
    \Gamma(\mathbf{r}_1,\mathbf{r}_2,\tau)=&\sum_{l=-\infty}^\infty\sum_{m=-\infty}^\infty \Gamma_{lm}(\pmb{\xi}_1,\pmb{\xi}_2,\tau)e^{i(m\phi_2-l\phi_1)}, \label{MCF}\\
    \Gamma_{lm}(\pmb{\xi}_1,\pmb{\xi}_2,\tau)=&\frac{1}{(2\pi)^2}\iint_0^{2\pi}\Gamma(\mathbf{r}_1,\mathbf{r}_2,\tau)e^{-i(m\phi_2-l\phi_1)}d\phi_1d\phi_2.\label{MCF-2}
\end{align}
In the space-frequency domain, the spatial coherence between the fields at $\mathbf{r}_1$ and $\mathbf{r}_2$ and at frequency $\omega$ is characterized by the cross-spectral density function (CSDF) $W(\mathbf{r}_1,\mathbf{r}_2,\omega)$, defined by the Wiener--Khintchine relations [28]
\begin{align}
    W(\mathbf{r}_1,\mathbf{r}_2,\omega)=&\frac{1}{2\pi}\int_{-\infty}^\infty \Gamma(\mathbf{r}_1,\mathbf{r}_2,\tau)e^{i\omega\tau}d\tau, \label{WK-1} \\ 
    \Gamma(\mathbf{r}_1,\mathbf{r}_2,\tau)=&\int_{0}^\infty W(\mathbf{r}_1,\mathbf{r}_2,\omega) e^{-i\omega\tau}d\omega. \label{WK-2}
\end{align}
Substituting Eq.~(\ref{WK-2}) to Eq.~(\ref{MCF-2}) leads to
\begin{align}\label{A5}
    \Gamma_{lm}(\pmb{\xi}_1,\pmb{\xi}_2,\tau)=\frac{1}{(2\pi)^2}\iint_0^{2\pi} \left[\int_0^\infty W(\mathbf{r}_1,\mathbf{r}_2,\omega)e^{-i\omega\tau}d\omega \right] e^{-i(m \phi_2-l \phi_1)}d\phi_1d\phi_2.
\end{align}
Similarly to the MCF in Eq.~(\ref{MCF}), the CSDF can be expressed as [31]
\begin{align}\label{W-coam}
    W(\mathbf{r}_1,\mathbf{r}_2,\omega)=\sum_{l=-\infty}^\infty\sum_{m=-\infty}^\infty W_{lm}(\pmb{\xi}_1,\pmb{\xi}_2,\omega)e^{i(m\phi_2-l\phi_1)},
\end{align}
where $W_{lm}(\pmb{\xi}_1,\pmb{\xi}_2,\omega)$ are the elements of the space-frequency domain COAM matrix. Combining Eqs.~(\ref{W-coam}) and (\ref{A5}), 
changing the order of integrations and summations, and employing the integral 
\begin{align}
\frac{1}{2\pi}\int_0^{2\pi}e^{i(\alpha-\beta)x}dx=\delta_{\alpha\beta},    
\end{align}
we find that 
\begin{align}\label{GammalmWlm}
    \Gamma_{lm}(\pmb{\xi}_1,\pmb{\xi}_2,\tau)=\int_0^\infty W_{lm}(\pmb{\xi}_1,\pmb{\xi}_2,\omega)e^{-i\omega\tau}d\omega.
\end{align}
The space-time and space-frequency domain COAM matrices thus satisfy a Wiener--Khintchine type relation in the polar Fourier space. 

\section{Derivation of the intensity at the detector in Fig.~1(\MakeLowercase{b}).}\label{App-meas}
In the following we derive the 
intensity at detector $\mathrm{D}_1$ in the setup depicted in Fig.~1(b). A similar result can be naturally established for the intensity at detector $\mathrm{D}_2$ by following the approach presented here. Assume that a quasi-monochromatic, paraxial field $U(\pmb{\rho},t)$, $\pmb{\rho}=[\rho,\phi]$ at a plane $z=0$ enters a combination of a spiral phase plate of order $-l$, a thin lens with focal length $f$, and a ring aperture of radius $\rho_0$ and width $\Delta$ placed directly behind the lens. The detector is at the distance $f$ from the system. The instantaneous intensity at the focal plane of the lens is obtained as (see Sect.~7.1 of [30])
\begin{align}
     I(\pmb{\rho},t)=\frac{1}{(\lambda f)^2}\iiiint T^\ast(\pmb{\rho}_1^\prime)T(\pmb{\rho}_2^\prime)J(\pmb{\rho}_1^\prime,\pmb{\rho}_2^\prime,t)\exp\left[-\frac{ik}{f}\pmb{\rho}\cdot(\pmb{\rho}_2^\prime-\pmb{\rho}_1^\prime)\right]d^2\pmb{\rho}_1^\prime d^2\pmb{\rho}_2^\prime,
\end{align}
where $\lambda$ is the central wavelength, $k=2\pi/\lambda$, $T(\pmb{\rho})$ is the combined transmission function of the ring aperture and spiral phase plate, and $J(\pmb{\rho}_1,\pmb{\rho}_2,t)=U^\ast(\pmb{\rho}_1,t)U(\pmb{\rho}_2,t)$ is the instantaneous mutual intensity of the beam at the input plane of the system. We further assume that the slit width $\Delta$ is narrow enough such that the transmission function can be approximated as $T(\pmb{\rho})=e^{-il\phi}\delta(\rho-\rho_0)\Delta$, where $\delta(x)$ is the Dirac delta. Then, evaluating the form above at the focal point $\rho=0$, we find that 
\begin{align}
    I(0,t)=\frac{4\pi^2\Delta^2\rho_0^2}{\lambda^2 f^2}\iint_0^{2\pi}J(\rho_0,\phi_1^\prime,\rho_0,\phi_2^\prime,t)e^{-il(\phi_2^\prime-\phi_1^\prime)}d\phi_1^\prime d\phi_2^\prime=\frac{4\pi^2\Delta^2\rho_0^2}{\lambda^2 f^2}I_l(\rho_0,t),
\end{align}
where the non-averaged version of Eq.~(\ref{MCF-2}) has been used to obtain the second equality. Thus, we see that measuring the instantaneous intensity at the focal spot of the lens gives the intensity $I_l(\rho_0,t)$ of the mode $U_l(\rho_0,t)$.

\section{Proofs related to the limits of Eq.~(19)}\label{appB}
In this section, we show that the limits $\bar{\gamma}(\pmb{\xi}_1,\pmb{\xi}_2,\tau)=1$ and $\bar{\gamma}(\pmb{\xi}_1,\pmb{\xi}_2,\tau)=0$ correspond to all pairs of modes $U_l(\pmb{\xi}_1,t)$, $U_m(\pmb{\xi}_2,t+\tau)$, being completely correlated and uncorrelated, respectively. We begin by introducing the correlation coefficients [31]
\begin{align}
    \gamma_{lm}(\pmb{\xi}_1,\pmb{\xi}_2,\tau)=\frac{\langle U_l^\ast(\mathbf{r}_1,t)U_m(\mathbf{r}_2,t+\tau)\rangle}{\sqrt{\langle |U_l(\pmb{\xi}_1,t)|^2\rangle \langle |U_m(\pmb{\xi}_2,t)|^2\rangle}},
\end{align}
which are limited as $0\leq |\gamma_{lm}(\pmb{\xi}_1,\pmb{\xi}_2,\tau)|\leq 1$, with the lower and upper bounds corresponding to $U_l(\pmb{\xi}_1,t)$ and $U_m(\pmb{\xi}_2,t+\tau)$ being completely uncorrelated and fully correlated, respectively. We may now write the (squared) OAMDC in the form 
\begin{align}
    &\bar{\gamma}^2(\pmb{\xi}_1,\pmb{\xi}_2,\tau)=\frac{\sum\limits_{l,m\in\mathcal{L}}|\gamma_{lm}(\pmb{\xi}_1,\pmb{\xi}_2,\tau)|^2 \langle |U_l(\pmb{\xi}_1,t)|^2\rangle \langle |U_m(\pmb{\xi}_2,t)|^2\rangle}{\sum\limits_{l,m\in\mathcal{L}} \langle |U_l(\pmb{\xi}_1,t)|^2\rangle \langle |U_m(\pmb{\xi}_2,t)|^2\rangle}.
\end{align}
From this expression it is clear that $\bar{\gamma}^2(\pmb{\xi}_1,\pmb{\xi}_2,\tau)=1$ if and only if $|\gamma_{lm}(\pmb{\xi}_1,\pmb{\xi}_2,\tau)|=1$ for all $(l,m)$, i.e., when all the components $U_l(\pmb{\xi}_1,t)$ and $U_m(\pmb{\xi}_2,t+\tau)$ are fully correlated. In addition, 
\begin{align}
&\mathrm{tr}[\overleftrightarrow{\Gamma}^\dagger(\pmb{\xi}_1,\pmb{\xi}_2,\tau)\overleftrightarrow{\Gamma}(\pmb{\xi}_1,\pmb{\xi}_2,\tau)]=\sum_{l,m\in\mathcal{L}}|\langle U_l^\ast(\pmb{\xi}_1,t)U_m(\pmb{\xi}_2,t+\tau)\rangle|^2,    
\end{align}
which attains the value of zero if and only if $\gamma_{lm}(\pmb{\xi}_1,\pmb{\xi}_2,\tau)=0$ for all $(l,m)$. Consequently, $\bar{\gamma}(\pmb{\xi}_1,\pmb{\xi}_2,\tau)=0$ corresponds to all the modes $U_l(\pmb{\xi}_1,t)$ and $U_m(\pmb{\xi}_2,t+\tau)$ being completely uncorrelated. 

\section{Derivation of Eq.~(23)}\label{Eq23proof}
In this section we show the derivation of Eq.~(23). From Eqs. (18) and (22) we see that
\begin{align}
    \bar{g}(\pmb{\xi}_1,\pmb{\xi}_2,\tau)=\frac{1}{L}|\bar{\gamma}_0(\pmb{\xi}_1,\pmb{\xi}_2,\tau)|^2+\frac{1}{2}\sum_{n=1}^{L^2-1} |\bar{\gamma}_n(\pmb{\xi}_1,\pmb{\xi}_2,\tau)|^2,
\end{align}
which, after taking the common factor and using Eq.~(21), can be written as
\begin{align}\label{S15}
    \bar{g}(\pmb{\xi}_1,\pmb{\xi}_2,\tau)=\frac{1}{L}|\bar{\gamma}_0(\pmb{\xi}_1,\pmb{\xi}_2,\tau)|^2\left[1+\frac{L}{2}\frac{\sum_{n=1}^{L^2-1} |\bar{\mathcal{S}}_n(\pmb{\xi}_1,\pmb{\xi}_2,\tau)|^2}{|\bar{\mathcal{S}}_0(\pmb{\xi}_1,\pmb{\xi}_2,\tau)|^2}\right].
\end{align}
Next, we define the quantity $\mathcal{Q}(\pmb{\xi}_1,\pmb{\xi}_2,\tau)$ as in Eq.~(24):
\begin{align}\label{S16}
    \mathcal{Q}(\pmb{\xi}_1,\pmb{\xi}_2,\tau)=\sqrt{\frac{L}{2(L-1)}\frac{\sum_{n=1}^{L^2-1} |\bar{\mathcal{S}}_n(\pmb{\xi}_1,\pmb{\xi}_2,\tau)|^2}{|\bar{\mathcal{S}}_0(\pmb{\xi}_1,\pmb{\xi}_2,\tau)|^2}}.
\end{align}
Substituting Eq.~(\ref{S16}) in Eq.~(\ref{S15}) immediately leads to
\begin{align}
        \bar{g}(\pmb{\xi}_1,\pmb{\xi}_2,\tau)=\frac{1}{L}|\bar{\gamma}_0(\pmb{\xi}_1,\pmb{\xi}_2,\tau)|^2\left[1+(L-1)\mathcal{Q}^2(\pmb{\xi}_1,\pmb{\xi}_2,\tau)\right],
\end{align}
which completes the derivation.

\section{Definition of $Q(\pmb{\xi})$ as a spectral distance}\label{appC}  

Here we show that $Q(\pmb{\xi})$ can be interpreted as the spectral distance of the OM from a uniform distribution. 
We begin by noting that the normalized eigenvalues $\sigma(\pmb{\xi})$ satisfy
\begin{equation}\label{muncond}
\sum\limits_{n=1}^{L}\sigma_n(\pmb{\xi})=1.
\end{equation}
Consequently, 
$0\leq \sigma_l(\pmb{\xi}) \leq 1$, 
defining a probability distribution, i.e., vector $\vec{\sigma}(\pmb{\xi})=[\sigma_l(\pmb{\xi})]$ lies on the simplex. We then consider the spectral Euclidean distance (variance) 
$\| \vec{\sigma}(\pmb{\xi})-\vec{I} \|_2$,
where $\vec{I}=\vec{1}/L$ and $\vec{1}$ is an $L$D vector of ones, normalized to its maximum value: \begin{equation}\label{M1}
\frac{\| \vec{\sigma}(\pmb{\xi})-\vec{I} \|_2}{\text{max}\| \vec{\sigma}(\pmb{\xi})-\vec{I} \|_2}.
\end{equation}

Since   
$\text{max} \| \vec{\sigma}(\pmb{\xi})-\vec{I} \|_2$ is achieved in the corners of the simplex, i.e., if $\sigma_1(\pmb{\xi})=1$ and $\sigma_l(\pmb{\xi})=0$, $l\geq 2$, we find that 
\begin{equation}
\begin{split}
\text{max}\| \vec{\sigma}(\pmb{\xi})-\vec{I} \|_2^2& =
\left( \frac{L-1}{L}\right)^2-\frac{L-1}{L^2}=\frac{L-1}{L}.
\end{split}
\end{equation}
Then, after opening the brackets and using Eq.~(\ref{muncond}), the numerator of Eq.~(\ref{M1}) becomes 
\begin{equation}
\| \vec{\sigma}(\pmb{\xi})-\vec{I} \|_2^2=\sum\limits_{l=1}^{L}\left[ \sigma_l(\pmb{\xi})-\frac{1}{L}\right]^2=\sum\limits_{l=1}^{L}\sigma_l^2(\pmb{\xi})-\frac{1}{L}. 
\end{equation}
Thus, as is evident from Eqs. (28) and \eqref{muncond}, the ratio in Eq. \eqref{M1} coincides with $Q(\pmb{\xi})$:
\begin{equation}\label{M2}
Q^2(\pmb{\xi})=\frac{L}{L-1}\left[ \sum\limits_{l=1}^{L}\sigma_l^2(\pmb{\xi})-\frac{1}{L}\right].
\end{equation}

\section{Derivation of Eq.~(29)}\label{appD}
Here we establish the COAM matrix displayed in Eq.~(29). The MCF for the incoherent superposition of LG and $I_l$-Bessel correlated modes is given by
\begin{align}
    \Gamma(\mathbf{r}_1,\mathbf{r}_2,\tau)=\Gamma_\mathrm{LG}(\mathbf{r}_1,\mathbf{r}_2,\tau)+\Gamma_\mathrm{B}(\mathbf{r}_1,\mathbf{r}_2,\tau).
\end{align}
Above, 
\begin{align}\label{GammaLG}
    \Gamma_\mathrm{LG}(\mathbf{r}_1,\mathbf{r}_2,\tau)=U_\mathrm{LG}^\ast(\mathbf{r}_1)U_\mathrm{LG}(\mathbf{r}_2)e^{-i\omega\tau}
\end{align}
is the portion of the MCF associated with the fully coherent LG modes, with $U_\mathrm{LG}(\mathbf{r})=\sum_{l\in\mathcal{L}}U_l(\pmb{\xi})e^{il\phi}$ and [18] 
\begin{align}
    U_l(\pmb{\xi})=& 
     \frac{w_0}{w(z)} \sqrt{\frac{2}{\pi|l|!}} \left[\frac{\rho\sqrt{2}}{w(z)}\right]^{|l|} \exp\left[- \frac{\rho^2}{w^2(z)} \right] 
    \exp\left[i(|l|+1)\Phi(z) - i\frac{\omega_0}{c} \frac{\rho^2}{2R(z)}\right]. \label{Ul} 
\end{align}
Here $w(z) = w_0\left[ 1+(z/z_{R})^2 \right]^{1/2}$, $w_0$ is the beam waist, $z_{R}=\pi w_0/\lambda_0$ is the Rayleigh range, $\lambda_0$ and $\omega_0$ are the central wavelength and frequency, respectively, $c$ denotes the speed of light, $R(z)=z[1+(z_{R}/z)^2]$ is the radius of  curvature, and $\Phi(z)=\mbox{arctan}(z/z_{R})$. 
Employing Eq.~(\ref{MCF-2}), the COAM matrix corresponding to $\Gamma_\mathrm{LG}(\mathbf{r}_1,\mathbf{r}_2,\tau)$ is found to be
\begin{align}
    \overleftrightarrow{\Gamma}_\mathrm{LG}(\pmb{\xi}_1,\pmb{\xi}_2,\tau)=\vec{U}^\dagger(\pmb{\xi}_1)\vec{U}(\pmb{\xi}_2)e^{-i\omega\tau},
\end{align}
where $\vec{U}(\pmb{\xi})=[U_l(\pmb{\xi})]$.
In addition, 
\begin{align}
    \Gamma_\mathrm{B}(\mathbf{r}_1,\mathbf{r}_2,\tau)=\int_0^\infty W_\mathrm{B}(\mathbf{r}_1,\mathbf{r}_2,\omega)e^{-i\omega\tau}d\omega
\end{align}
represents the MCF contribution from the $I_l$-Bessel correlated part of the beam, with the related CSDF taken to be
\begin{align}
    W_\mathrm{B}(\mathbf{r}_1,\mathbf{r}_2,\omega)
    =\sum_{l\in\mathcal{L}}W_l(\pmb{\xi}_1,\pmb{\xi}_2,\omega)e^{il(\phi_2-\phi_1)},
\end{align}
where [49]
\begin{align}
    W_l(\pmb{\xi}_1,\pmb{\xi}_2,\omega)=&W_f(\omega)\frac{\zeta^{-l/2}}{1-\zeta}\frac{w_0^2}{w(z_1)w(z_2)}
      \exp\left(i\left\{\frac{\omega}{c}\left(z_1-z_2\right)-(l+1)[\Phi(z_1)-\Phi(z_2)]\right\}\right)
      \nonumber \\&\times
     \exp\left\{i\frac{\omega}{c}\left[\frac{\rho_1^2}{2R(z_1)}-\frac{\rho_2^2}{2R(z_2)}\right]\right\}
         \exp\left\{-\frac{1+\zeta}{1-\zeta}\left[\frac{\rho_1^2}{w^2(z_1)}+\frac{\rho_2^2}{w^2(z_2)} \right]\right\}\nonumber\\
     &\times I_l\left[\frac{4\zeta^{1/2}}{1-\zeta}\frac{\rho_1\rho_2}{w(z_1)w(z_2)} \right]. 
\end{align}
Here $W_f(\omega)$ is a spectral weight factor, $I_l(x)$ is the modified Bessel function of order $l$, $\zeta$, $0<\zeta <1$, is a coherence parameter associated with the modes, with larger $\zeta$ corresponding to more incoherent light, and the other parameters are same as above. In particular, we may choose the spectral factor to be a Gaussian function
\begin{align}\label{Wf-Gauss}
    W_f(\omega)=
    \frac{1}{\delta_\omega\sqrt{2\pi}}\exp\left[-\frac{(\omega-\omega_0)^2}{2\delta_\omega^2}\right],
\end{align}
where 
$\delta_\omega$ 
is spectral bandwidth. The components $W_l(\pmb{\xi}_1,\pmb{\xi}_2,\omega)$ represent the diagonal COAM matrix elements in the space-frequency domain, with the off-diagonals taken to be zero. We then obtain by evaluating Eq.~(\ref{GammalmWlm}) the diagonal elements of the space-time domain COAM matrix at a plane of constant $z$:
\begin{align}\label{GammaB}
    \Gamma_{ll}^{(\mathrm{B})}(\rho_1,\rho_2,z,\tau)=&\frac{\zeta^{-l/2}}{1-\zeta}\left[\frac{w_0}{w(z)}\right]^2
    \exp\left\{i\left[\frac{\omega_0}{c}\frac{\rho_1^2-\rho_2^2}{2R(z)}-\omega_0\tau\right]\right\}
    \exp\left\{-\frac{\delta_\omega^2}{2}\left[\frac{1}{c}\frac{\rho_1^2-\rho_2^2}{2R(z)}-\tau\right]^2\right\}
    \nonumber\\&\times
    \exp\left\{-\frac{1+\zeta}{1-\zeta}\left[\frac{\rho_1^2+\rho_2^2}{w^2(z)}\right]\right\}
     I_l\left[\frac{4\zeta^{1/2}}{1-\zeta}\frac{\rho_1\rho_2}{w^2(z)} \right].
\end{align}
Combining Eqs.~(\ref{GammaLG}) and (\ref{GammaB}) and setting $\tau=0$, we obtain Eq.~(29):
\begin{align}
    \overleftrightarrow{\Gamma}(\rho_1,\rho_2,z)=\vec{U}^\dagger(\rho_1,z)\vec{U}(\rho_2,z)+\overleftrightarrow{R}(\rho_1,\rho_2,z),
\end{align}
where $\overleftrightarrow{R}(\rho_1,\rho_2,z)=\mathrm{diag}[R_l(\rho_1,\rho_2,z)]$ and
\begin{align}
    R_l(\rho_1,\rho_2,z)=&\frac{\zeta^{-l/2}}{1-\zeta}\left[\frac{w_0}{w(z)}\right]^2
    \exp\left[i\frac{\omega_0}{c}\frac{\rho_1^2-\rho_2^2}{2R(z)}\right]
    \exp\left[-\frac{\delta_\omega^2}{8c}\frac{(\rho_1^2-\rho_2^2)^2}{R^2(z)}\right]
    \nonumber\\&\times
    \exp\left\{-\frac{1+\zeta}{1-\zeta}\left[\frac{\rho_1^2+\rho_2^2}{w^2(z)}\right]\right\}
     I_l\left[\frac{4\zeta^{1/2}}{1-\zeta}\frac{\rho_1\rho_2}{w^2(z)} \right].
\end{align}